\documentclass[aps,pre,onecolumn,showpacs,a4paper]{revtex4}
\usepackage{graphicx} 
\usepackage{amsmath}
\usepackage{amssymb}
\usepackage{amscd}

\begin{document}

\title{Anomalous diffusion in a random
 nonlinear oscillator due to  high frequencies of the noise}

\author{K. Mallick}
 \affiliation{Service de Physique Th\'eorique, CEA Saclay, 91191
  Gif-sur-Yvette, France}
\date{\today}

\begin{abstract}
  We study the long time behaviour of a   nonlinear oscillator subject
 to a random multiplicative  noise with  a spectral density  (or
 power-spectrum)  that decays as a  power law at high frequencies.
 When the dissipation is negligible, physical observables, such as the
 amplitude, the velocity and the energy of the oscillator grow  as
 power-laws with time.  We calculate the associated  scaling exponents
 and we show that their values  depend  on the asymptotic behaviour of
 the   external potential  and on the  high frequencies of the noise.
 Our results are generalized to include dissipative effects   and
 additive noise.
  \end{abstract}
 \pacs{05.10.Gg,05.40.-a,05.45.-a}
\maketitle

   The interplay of randomness  with nonlinearity gives rise to a
  variety of phenomena both of practical and 
  theoretical significance that have been the subject of
   many studies  since the pioneering work  of 
   Stratonovich  \cite{strato,lefever,anishchenko,pikovsky}
   For example,  electronic and  Josephson
  junctions  subject to thermal noise \cite{kabashima,devoret}, 
   the Faraday instability
  of surface waves  \cite{petrelis}  and even human rythmic movements
   \cite{daffert}  can be
   modelized by nonlinear oscillators  subject  to various
    sources of noise (for a general reference on this subject, see e.g.,
    \cite{gitterman}).
   In the regime where   the dissipation
  of the effective nonlinear oscillator  can be  neglected,
 the random perturbation  injects energy into
   the system and  physical observables
  such as  the oscillator's mechanical energy, root-mean-square
  position and velocity, grow algebraically with time
 \cite{bourret,schentzle,landaMc}.  This instability saturates
 when  dissipative effects (that scale  typically as the square-root
 of the energy) become  large enough to overcome  the injection rate and  
  the system   reaches an non-equilibrium steady state. 
  In series of recent  papers \cite{philkir1,philkir2,philkir3},
  we have derived analytical results
 for quasi-Hamiltonian nonlinear random oscillators
  subject to external or internal random perturbation  that is either
 a Gaussian white noise or an Ornstein-Uhlenbeck process. 
 In particular, we have calculated the
  scaling exponents characterize the algebraic growth 
   of the energy, the position 
   and the velocity of the oscillator.

  In the present work,  we consider the general case  a nonlinear oscillator 
  subject to a  multiplicative
   noise with an   arbitrary  correlation function.
  We  show  that the physical observables display a scaling behaviour
 in the long time limit and we 
  calculate the  associated  scaling exponents. 
  The formulae obtained  
  are valid for any type  of noise whose power-spectrum decays as a power-law
 at large frequencies; in particular,
 the results previously derived for white and Ornstein-Uhlenbeck
 noises  appear as  special cases of the general expressions that
 are derived in the present work.  Finally, we 
   briefly explain how our results can be
 generalized to include dissipative effects,  additive noise,
 and how the scaling behaviour is modified when the noise 
 power-spectrum decays decays exponentially at large frequencies.

\section{The dynamical equations in action-angle variables}

  We  consider a   nondissipative
  oscillator of amplitude $x(t)$, trapped
 in  a  nonlinear  confining potential ${\mathcal U}(x)$
   and subject to a multiplicative
 noise $\xi(t)$~:
 \begin{equation}
   \frac{\textrm{d}^2 }{\textrm{d} t^2}x(t) 
   = - \frac { \partial{\mathcal U}(x)}{\partial x}  + x(t)\, \xi(t) \,.
 \label{dynamique}
\end{equation}
   Our aim is to  study the effect of
  the  statistical properties of  $\xi(t)$ on the long
 time properties of the dynamical variable $x(t)$.

 We restrict our analysis to the case where the dominant
 term in the  potential  ${\mathcal U}$  when 
 $ |x| \rightarrow \infty$ is  an even power of  $x$.
   A suitable rescaling
  of $x$ allows us to write
 \begin{equation}
     {\mathcal U}   \sim \frac{ x^{2n}}{2n}
  \,\, \hbox{ with } \,\, n \ge  2 \,. 
 \label{infU}
\end{equation}
As the amplitude $x(t)$ of the oscillator grows with time,
  the  leading  behavior of ${\mathcal U}(x)$ for 
$ |x| \rightarrow \infty$  only is  relevant and Eq.(\ref{dynamique})
 reduces to
  \begin{equation}
   \frac{\textrm{d}^2 }{\textrm{d} t^2}x(t) 
 + x(t)^{2n-1}  = x(t) \, \xi(t)  \,.
 \label{dyn2}
\end{equation}

 We shall  analyse the motion of the  nonlinear stochastic oscillator
    following the method
 explained in \cite{philkir1}. Defining the energy
 and  the angle variables, 
\begin{equation}
  E =  \frac{1}{2}\dot x^2 + \frac{1}{2n} x^{2n} \, \,\, ,
  \;\;\; \hbox{ and } \;\;\;   \phi  = 
 \frac{ \sqrt{n}} { (2n)^{1/2n} } \int_0^ { {x}/{E^{{1}/{2n}}} }
   \frac{{\textrm d}u}{\sqrt{ 1 -  \frac{u^{2n}}{2n}}}    \, , 
\label{defphi}
\end{equation}
 we  transform  the coordinates in phase space 
 from position and velocity  to   energy and angle: 
\begin{eqnarray}
          x &=&   E^{\frac{1}{2n}} \, { s}_n
 \left( \phi  \right) , \label{solnxv2}\\                 
     \dot x &=& (2n)^{\frac{n-1}{2n}} E^{ \frac{1}{2} } \,
  { s}_n'\left( \phi  \right) \, ,
\label{solnxv3}
 \end{eqnarray}
where  the hyperelliptic function ${s}_n$ is defined by the relation 
 ${s}_n(\phi) = x/ E^{\frac{1}{2n}}$.
  Using  the   auxiliary variable $\Omega$, that satisfies
  \begin{equation}
  \Omega =   (2n)^{ \frac{n+1}{2n} } \,  E^{\frac{n-1}{2n}}  \, , 
 \label{defOmega}
  \end{equation}
 equation~(\ref{dyn2}) is  rewritten  \cite{philkir1,philkir2}
   as a system of two coupled
 stochastic differential equations 
   \begin{eqnarray}
     \dot \Omega  &=& (n -1) \, { s}_n(\phi)
 { s}_n'(\phi) \, \xi(t)  
   \,  ,      \label{evolomega} \\
 \dot\phi  &=& \frac{\Omega}{ (2n)^{\frac{1}{n}}}
 - \frac{{s}_n(\phi)^2}{\Omega}  \, \xi(t)  \, .
    \label{evolphi}
   \end{eqnarray}
 This system is  rigorously 
 equivalent to the original  problem   (Eq.~\ref{dyn2}) and has been  
  derived without any hypothesis   on the 
 random perturbation term $\xi(t)$. 
  It  appears clearly in  this formulation  that the angle $\phi$
 is a fast variable whereas the action $\Omega$ is a slow variable~: 
  an effective stochastic equation for $\Omega$ can thus be 
  derived  by adiabatic averaging over the angular variable. 

 We now specify the  statistical properties of the 
 random perturbation   $\xi(t)$.
 We shall  consider  a stationary  Gaussian noise
  of zero mean value. A  Gaussian process is  entirely
 characterized by its auto-correlation function, defined as
 \begin{equation}
  {\mathcal S}(t'-t) = \langle \xi(t')  \xi(t) \rangle  \, .
\label{eq:defS}
 \end{equation}
  In Fourier  space,   the power-spectrum  of the noise is given by 
  \begin{equation}
  \hat{\mathcal S}(\omega) = \int_{-\infty}^{+\infty} {\rm d}t 
        \exp(i\omega t)   {\mathcal S}(t) = 
  \int_{-\infty}^{+\infty} {\rm d}t \exp(i\omega t)
    \langle  \xi(t)  \xi(0)  \rangle  \, .
\label{eq:PSD}
 \end{equation}
  For instance, if  $\xi(t)$ is  a  white noise
 of  amplitude ${\mathcal D}$,  we have 
 $\hat{\mathcal S}(\omega) =   {\mathcal D}$. 
 When $\xi(t)$ is an Ornstein-Uhlenbeck process of amplitude
  ${\mathcal D}$ and autocorrelation time $\tau$,   we have 
\begin{equation}
  {\mathcal S}(t'-t) =   
\frac{ {\mathcal D} }{2 \, \tau}   \, {\rm e}^{-|t - t'|/\tau} \quad 
   \hbox{ and } \quad  \hat{\mathcal S}(\omega) =  
   \frac{ {\mathcal D} }{ 1 + \omega^2\tau^2} \, . 
   \label{deftau}
 \end{equation}  
 In the white noise case, we derived  in \cite{philkir1} 
   the  following scaling relations, valid when $t \to \infty$
\begin{equation}
          E      \sim  \left({\mathcal D}t\right)^{\frac{n}{n-1}} , 
  \quad
          x     \sim   \left({\mathcal D}t\right)^{\frac{1}{2(n-1)}} ,
 \quad
         \dot x   \sim   \left({\mathcal D}t\right)^{ \frac{n}{2(n-1)}}  \, .
\label{scalingwhite}
\end{equation}
  In the case when  $\xi(t)$ is an Ornstein-Uhlenbeck process,  
 we found in  \cite{philkir2} that, in the long time limit,
 the energy, the amplitude and the velocity of the oscillator grow as~:
\begin{equation}
   E  \sim
  \left(\frac{ {\mathcal D}t} { \tau^2} \right)^{\frac{n}{2(n-1)}} , 
  \quad
 x   \sim  
  \left(\frac{ {\mathcal D}t} { \tau^2} \right)^{\frac{1}{4(n-1)}} ,
    \quad
 \dot x  \sim  
  \left(\frac{{\mathcal D}t} { \tau^2}\right)^{\frac{n}{4(n-1)}} \, . 
\label{scalingOU}
\end{equation} 
   These  colored noise scalings
 are  always observed when $t \to \infty$, even if 
 the correlation time $\tau$ is arbitrarily small. 
 We remark in equations~(\ref{scalingwhite}) and~(\ref{scalingOU})
  that the  scaling  exponents only depend on the asymptotic
 growth rate of the potential. However, the exponents 
 are reduced by a factor 2 for  Ornstein-Uhlenbeck
 noise as compared to their value for 
  white noise; this means  physically that
 the energy transfer from the noise to the oscillator
 is less efficient when the noise is correlated in time.  
  In  \cite{philkir2}, we derived an analytic expression
  for the  probability distribution function (PDF) in the long time
 limit that allowed us to derive not only the scaling laws but
 also to calculate all the prefactors. However, we did not
 find any satisfactory explanation  that would
 explain the origin  of the factor 2  reduction in 
 of the scaling exponents.  We  show in the next section by analyzing
  the general case that the scaling exponents not only depend
 on the potential at infinity but also on the asymptotic behaviour
 of the spectral density of the noise at high frequencies. 


\section{Scaling behaviour in presence of arbitrary noise}

 We consider now  the more  general case  of a Gaussian
  noise   which can be generated from  the  white noise 
 by solving a linear differential equation of order $\sigma$.
  Such a noise has a  power-spectrum
  that  decays as follows   at high frequencies~:
 \begin{equation}
      \hat{\mathcal S}(\omega)  \sim 
    {\mathcal D} (\omega\tau)^{-2\sigma}
   \,\,\,\, \hbox{ when } \,\,\,\,\,  |\omega| \rightarrow \infty \, .
\label{eq:defsigma}
  \end{equation}
  The amplitude  ${\mathcal D}$ of the noise
  and  the correlation-time $\tau$ are defined
   by dimensional analogy with equation~(\ref{deftau}). The 
 exponent $\sigma$ characterizes the high frequency behaviour  of the 
 the  power-spectrum of the noise.   We shall show in this section
 that,  in the 
 long time limit, the scaling behaviour of the nonlinear oscillator
  in presence of a noise with a  power-spectrum
  that satisfies equation~(\ref{eq:defsigma}), is given by
  \begin{eqnarray}
     E  &\sim&
    \left(\frac{ {\mathcal D}t} {\tau^{2\sigma}}
   \right)^{\frac{n}{(\sigma+1)(n-1)}}  \,  , 
   \nonumber   \\
 x  &\sim&  
  \left(\frac{ {\mathcal D}t} { \tau^{2\sigma}}
    \right)^{\frac{1}{2(\sigma+1) (n-1)}} \, ,
   \nonumber    \\ 
 \dot x &\sim& 
  \left(\frac{{\mathcal D}t} { \tau^{2\sigma}} 
  \right)^{\frac{n}{2(\sigma+1) (n-1)}} \,  . 
\label{scalinggeneral}
\end{eqnarray}
 For $\sigma =0$ (respectively  $\sigma =1$) 
 we recover  the  scalings given in equation~(\ref{scalingwhite})
 for a white noise  (respectively the  scalings 
   given in  equation~(\ref{scalingOU}) for 
  Ornstein-Uhlenbeck noise). The formulae~(\ref{scalinggeneral})
  provide us with the explicit dependence of the growth exponents
  on  the confining potential and  on the noise spectral density.
  In particular,  we observe that when the relative weight
 of the  large frequencies
 is  decreased  (by increasing  the value of  $\sigma$)
 the effective  diffusion in the oscillator phase space  becomes slower.

   We now  derive  the power laws~(\ref{scalinggeneral}). 
  One method to obtain these scalings is to apply the 
   recursive adiabatic averaging  procedure
   developped  in \cite{philkir2,philkir3}.
    This advantage of this technique is to provide explicit 
   expressions for the prefactors in  the scaling law.
  The basic idea is  the following~: the naive averaging
 procedure applied to the 
  system~(\ref{evolomega},\ref{evolphi})
   fails if the correlations
 between the angle $\phi$ and the noise  $\xi(t)$  are discarded.
 However, by taking the  derivatives
   of the  equations~(\ref{evolomega},\ref{evolphi})
  repeatedly,  a white noise contribution will appear 
 (after $\sigma$ derivatives) and will provide  the leading  
  term in long-time behaviour of the  oscillator. However, 
 in order to apply this procedure, we need an explicit 
 differential equation that relates the noise $\xi(t)$
 to the white noise. Furthermore, this technique,
  though  quantitatively precise, leads to 
  very intricate  calculations and is suitable for simple noises
  such as the  Ornstein-Uhlenbeck process  or the harmonic noise.

  We shall therefore  derive here the  power  laws~(\ref{scalinggeneral})
  from a  simpler argument. First, 
    because  we only want to extract the  leading scaling behaviour
   when  $t \to \infty$, 
  we simplify  the
  equations~(\ref{evolomega}) and~(\ref{evolphi}), 
   as we did in \cite{philkir2}~: we 
   keep  only the dominant terms, replace  the 
    hyperelliptic function ${s}_n(\phi)$
  by  the  circular  function $\sin\phi$ and 
  put  all the numerical factors to 1.   We  thus 
  obtain  the following system
    \begin{equation}
     \dot \Omega  =  \, \xi(t)  \sin\phi   
    \quad \hbox{ and } \quad 
 \dot\phi  =  \Omega  \, , 
    \label{pendulum}
   \end{equation}
 which describes a pendulum with random frequency \cite{philkir4}.
  Now,   the second order cumulant
 expansion~\cite{vankampen} of the stochastic Liouville equation
 for the PDF  $P_t(\Omega,\phi)$
  associated with equations~(\ref{pendulum})  is given by 
\begin{equation}
\frac{ \partial   P_t }{\partial t}  = 
 {\bf L}_0 P_t  + \int_0^t d\theta \langle {\bf L}_1(t) \exp({\bf L}_0\theta)
    {\bf L}_1(t-\theta)   \exp(-{\bf L}_0\theta) \rangle P_t   
\label{FPcumul}
\end{equation}
where  the differential operators are defined as 
 \begin{eqnarray}
    {\bf L}_0  P_t &=&  - \frac{\partial}{\partial \phi}( \Omega P_t )  \, ,
\label{defL0}  \\
    {\bf L}_1(t) P_t &=&  - \frac{ \partial}{\partial \Omega}
 ( \xi(t) \sin\phi \,  P_t) \,.\label{defL1}
  \end{eqnarray}
 Following the method developed in    \cite{lindcol},
 we  evaluate  the right hand side of equation~(\ref{FPcumul}) by using 
 to the formula 
 \begin{equation}
 \exp(A) B \exp(-A) =  B + [A,B] + \frac{1}{2!} [A, [A,B]] + 
 \frac{1}{3!} [A, [A, [A,B]]] + \ldots  \,, 
\label{idmat}
\end{equation}
 with  $ A = {\bf L}_0$   and $B = {\bf L}_1(t-\theta)$. We find by 
 induction  the expression of  the $n$-th  commutator 
\begin{equation}
[{\bf L}_0,[\ldots,[{\bf L}_0, [{\bf L}_0, {\bf L}_1(t-\theta)]]\ldots] 
  =  \xi(t-\theta)
  \left( \frac{ \partial   }{\partial \Omega}  H_1^{(n)}(\Omega,\phi)
   +  \frac{ \partial   }
 {\partial \phi}  H_2^{(n)}(\Omega,\phi)    \right)
 \, , \label{defTn}
\end{equation}
where the  functions $H_1^{(n)}$   and $H_2^{(n)}$ are given by 
 \begin{eqnarray}
   H_1^{(n)}(\Omega,\phi) &=& 
   (-1)^{n-1} \Omega^n \sin(\phi + n \frac{\pi}{2})  
     \, ,  \\
   H_2^{(n)}(\Omega,\phi) &=&   
 (-1)^{n-1} n  \Omega^{n-1} \cos(\phi + n \frac{\pi}{2})  
  \, . 
\label{solrecurrence}
\end{eqnarray}
 Substituting the equations~(\ref{idmat}-\ref{solrecurrence})
 in equation~(\ref{FPcumul}), we obtain, after 
 taking the average over the noise
  and integrating over the variable $\theta$~:
 \begin{equation}
 \int_0^t d\theta \langle {\bf L}_1(t) \exp({\bf L}_0\theta)
    {\bf L}_1(t-\theta)   \exp(-{\bf L}_0\theta) \rangle  P_t   = - 
   \frac{\partial}{\partial \Omega} \sin\phi  \,
 \left(  \frac{\partial}{\partial \Omega} {\mathcal H}_1(\Omega,\phi,t) P_t 
     +  \frac{ \partial}{\partial \phi}
    {\mathcal H}_2(\Omega,\phi,t)P_t  \right)    \, ,
\label{opdiff}
 \end{equation}
   with 
\begin{eqnarray} 
{\mathcal H}_1(\Omega,\phi,t)   &=& 
  \sum_{n=0}^{\infty}  \frac{ H_1^{(n)}(\Omega,\phi) }{n!}  \int_0^t 
  {\rm d}\theta  \,  \theta^n  {\mathcal S}(\theta) 
 =  -  \int_0^t \sin(\Omega \theta +\phi) {\mathcal S}(\theta){\rm d}\theta
  \, ,
  \label{defgotH1} \\
   {\mathcal H}_2(\Omega,\phi,t)  &=& 
  \sum_{n=0}^{\infty}  \frac{ H_2^{(n)}(\Omega,\phi) }{n!}  \int_0^t 
  {\rm d}\theta  \,  \theta^n  {\mathcal S}(\theta) 
 =  \int_0^t  \sin(\Omega \theta -\phi)
   \theta {\mathcal S}(\theta){\rm d}\theta  \, .  \label{defgotH2}
\end{eqnarray}
 In the limit $t \to \infty$,  we average
  the differential  operator~(\ref{opdiff}) over the fast variable $\phi$
 and we deduce  that 
 \begin{eqnarray}
 \int \frac{d\phi}{2\pi}
  \int_0^\infty d\theta \langle {\bf L}_1(t) \exp({\bf L}_0\theta)
    {\bf L}_1(t-\theta)   \exp(-{\bf L}_0\theta) \rangle 
 &=& \frac{1}{2} \frac{\partial^2}{\partial \Omega^2}\int_0^\infty
  \cos(\Omega \theta) {\mathcal S}(\theta){\rm d}\theta 
  -\frac{1}{2}   \frac{\partial}{\partial \Omega}
    \int_0^\infty  \sin(\Omega \theta)
   \theta {\mathcal S}(\theta){\rm d}\theta  \nonumber \\
  &=&  \frac{1}{4} \frac{\partial^2}{\partial \Omega^2}
   \hat{\mathcal S}(\Omega) -  \frac{1}{4} \frac{\partial}{\partial \Omega}
     \hat{\mathcal S}'(\Omega) \, , 
 \end{eqnarray}
 where  $\hat{\mathcal S}'(\Omega)$ represents the derivative
 of the power-spectrum with respect to  the frequency.  
  Finally,  we apply this averaged operator to  ${\tilde P}_t(\Omega)$,
  the  probability  distribution  function of 
 the slow variable $\Omega$ and we derive,
  in the long time limit, an 
  effective averaged Fokker-Planck equation for  ${\tilde P}_t(\Omega)$~:
\begin{equation}
  \frac{ \partial   {\tilde P}_t  } {\partial t}  = 
  \frac{1}{4}
    \frac{\partial}{\partial \Omega} \Big(  
     \hat{\mathcal S}(\Omega)
    \frac{\partial  {\tilde P}_t    }{\partial \Omega}  \Big)  \, .
   \label{effectiveFP}
\end{equation}
  This equation only involves the power spectrum of the noise. 
  If we substitute  in this equation 
 the large frequency behaviour~(\ref{eq:defsigma}) of $\hat{\mathcal S}$,
 we  conclude  by elementary dimensional analysis that~:
\begin{equation}
     \Omega^{2(\sigma +1)} \sim 
  \frac{ {\mathcal D}t} {\tau^{2\sigma}}  \, .
  \label{eq:scalOmega}
\end{equation}
   By  using the definition~(\ref{defOmega}) of $\Omega$ and  the 
 relations~(\ref{solnxv2}) and~(\ref{solnxv3}) between  the amplitude $x$, 
 the velocity $\dot{x}$  and  the energy $E$,  the scaling
  relations~(\ref{scalinggeneral}) are derived.

\begin{figure}
\label{fig1}
\end{figure}

    Our calculation can be generalized to the case
 of a weakly dissipative system, by including  a linear
  friction term in equation~(\ref{dynamique}) with 
  dissipation rate $\gamma$.  In this case, the oscillator's amplitude
 does not grow without bounds but saturates after a time
 of the order $1/\gamma$. The system reaches a non-equilibrium
 steady state described by a stationary probability,
 which is in general not a Gibbs-Boltzmann distribution.
 In the case of a vanishingly
 small rate, the saturation time is large and 
 the averaging technique can be still be applied. It can be shown, using
 the same method as above, that the stationary probability  behaves as 
 \begin{equation}
 P_{\hbox{stat}}(E) \sim
  \exp\Big( -C\frac{\gamma \tau^{2\sigma} E^{\frac{(\sigma+1)(n-1)}{n}}}
 { {\mathcal D} }
 \Big) \quad \hbox{ when } E \to \infty \, ,
\end{equation}
 where $C$ is a constant. The  analytical expression  of $C$
 and of the prefactors  in $P_{\hbox{stat}}$ can not be obtained
 by the method used here (a  recursive adiabatic averaging
  similar to the one used for white and Ornstein-Uhlenbeck noise
  would be needed to go beyond scaling). 

    We have asuumed hitherto   that the noise $\xi(t)$ is generated from
 white noise by solving a linear differential equation
 of order $\sigma$. Therefore,  $\xi(t)$ is Gaussian 
 and its   power-spectrum
 $\hat{\mathcal S}(\omega)$ decreases as a power-law at larges frequencies. 
   However,  the derivation of the
  effective Fokker-Planck equation~(\ref{effectiveFP})   does not 
  rely  on these assumptions and can still  be  performed when 
  the correlation function  of $\xi(t)$ has long-tails,
  for example, when 
  $\hat{\mathcal S}(\omega) \sim \exp( -\tau|\omega|) $
  (Lorentzian  power-spectrum). We then obtain that
  \begin{equation}
   \Omega \sim \frac{1}{\tau} \log\frac{t}{\tau} \, , 
 \end{equation}
  {\it i.e.},  the amplitude of the system grows logarithmically 
  with time.   Similarly,  if the power-spectrum of the
 noise is Gaussian, {\it i.e.}, $\hat{\mathcal S}(\omega) \sim
  \exp( -(\tau\omega)^2)$ at high frequencies, the action variable
 $\Omega^2$ - increases  logarithmically with time. 
   If the power-spectrum
  has a high-frequency cut-off at  $\omega_0$, 
 {\it i.e.}, if $\hat{\mathcal S}(\omega) = 0 $ for $\omega > \omega_0$
   we conjecture  $\Omega$  saturates  at large
 times. This fact is true,  in particular,  when 
 the noise is a deterministic circular function
  for example in the case  $\xi(t) = \cos(\omega_0 t)$.

   Finally, we  emphasize  that
  the method we have used here can also be applied to a
 nonlinear oscillator subject to an additive noise,
 described by the following  Langevin equation
\begin{equation}
   \frac{\textrm{d}^2 }{\textrm{d} t^2}x(t) 
   = - \frac { \partial{\mathcal U}(x)}{\partial x}  +  \xi(t) \, , 
 \label{dynamiqueadd}
\end{equation}
 where the  power-spectrum of the noise $\xi(t)$ satisfies
 the  equation~(\ref{eq:defsigma}). Then,  by calculations  similar
 to those described above, the following scaling behaviour
 is derived (an alternative method that leads to the same
 results would be to use
 the Markovian approximation for the energy dynamics in the low
 friction limit, developed in \cite{nitzan,schimansky})~:
\begin{equation}
   E  \sim
  \left(\frac{ {\mathcal D}t} 
   { \tau^{2\sigma}} \right)^{\frac{n}{\sigma(n-1) + n}} , 
  \quad
 x   \sim  
  \left(\frac{ {\mathcal D}t}
   { \tau^{2\sigma}} \right)^{\frac{1}{2\sigma(n-1) + 2n}} ,
    \quad
 \dot x  \sim  
  \left(\frac{{\mathcal D}t}
  { \tau^{2\sigma}}\right)^{\frac{n}{2\sigma(n-1) + 2n}} \, . 
\label{scalingadd}
\end{equation}
   In particular, when the noise is white, we recover the fact
 that the energy grows linearly with time. 

\section{Concluding remarks}

    A particle trapped in a nonlinear confining potential and subject
 to a random noise undergoes anomalous diffusion in phase space.
  In this work, we have found analytical expressions for 
  the associated  diffusion exponents, by using an 
   action-angle representation of the equations of motion that
 allowed  us  to separate  the slow variable from the fast variable
   and to derive an effective Langevin dynamics for the slow variable.
   Our results are derived for  an arbitrary correlation function of the noise
 and are valid for multiplicative noise as well as for additive noise.
  We  have shown
 that the high frequency components in the noise  spectrum
  play a crucial role~: the fastest the power spectrum  of the noise decays
  at  high frequencies, the slower is the diffusion. This fact 
 has the following physical interpretation~:  the energy transfer
 from  the external  driving to  the oscillator
 is optimal when the driving frequency is of the order of  the natural 
  frequency of the oscillator  ({\it e.g.},  parametric resonance
 is maximal for driving at twice the natural frequency). 
  However,  because  of the
 nonlinearity, the period $T$ of the underlying
 deterministic oscillator decreases with the  amplitude.  Therefore,
 at large amplitudes, the energy transfer between the noise and
 the system involves higher and higher frequencies. If these frequencies
 have a small weight in the noise  power spectrum, the amplification
 of the oscillator is less  efficient. Our calculations put this
 intuitive reasoning on a quantitative basis and can be generalized
 to include dissipative effects. In presence of dissipation, the
  system reaches a nonequilibrium steady state in  which
 the stationary PDF  is not the canonical Gibbs-Boltzmann
 distribution but  a streched exponential.

\subsection*{Acknowledgments}

   I am  thankful to  E. Bogomolny'i for a question that 
 was at the origin of this work.

\end{document}